\documentclass[12pt,preprint]{aastex}
\usepackage{epsfig}

\def\refe{}

\begin{document}
\title{Planets Rapidly Create Holes  in Young Circumstellar Discs}

\author{P.Varni\`ere$^{1}$, Eric.G. Blackman$^1$ A. Frank$^1$ \& Alice. C. Quillen$^1$}
\affil{1. Department of Physics and Astronomy, University of Rochester,
Rochester NY 14627, USA}

\begin{abstract}
Recent spectral observations by the Spitzer Space Telescope (SST) reveal that 
some discs around young ($\sim {\rm few} \times 10^6$ yr old) stars have 
remarkably sharp transitions  to a low density inner region
in  which much of the material has been cleared 
away.
It has been recognized that 
the most plausible mechanism for the sharp transition at a specific
radius is the gravitational influence of a massive planet. 
This raises the question of whether the planet can also account
for the hole extending all the way to the star.   
Using high resolution numerical simulations, we show that Jupiter-mass planets drive spiral waves 
   which create holes on time scales $\sim 10$ times shorter than viscous or planet 
migration times. {\refe We find that the theory of 
spiral-wave driven accretion in viscous flows by Takeuchi et al. (1996)
can be used to provide a consistent interpretation of the simulations.}
In addition, although the  hole surface  
densities are low, they are finite, allowing  mass accretion toward the star.
Our results therefore imply that massive planets  can form
extended, sharply bounded spectral holes which can still accommodate 
substantial mass accretion rates.
The results also imply
that holes are more likely than gaps for Jupiter mass planets
around solar mass stars.
\end{abstract}

\section{Introduction}

The discovery of extrasolar planets some $10$ years ago has led to a renaissance in our understanding 
of the formation and evolution of planetary systems \citep{MB98}. 
Planets form accretion disks 
that surround and feed mass onto young stars, but there is significant gravitational feedback 
between the planets and their parent disks \citep{L03}. This 
produces a variety of disk   architectures 
that often differ from our own solar system. Understanding planet-disk interactions is essential for 
understanding planetary positions, 
observational signatures for exo-planets, and the time scales 
that planet formation models must accommodate.

Despite significant theoretical progress \citep{L03}, key questions remain. The time 
scale for planet formation, $\tau_f$, remains uncertain.  Gas giant planets might form rapidly ($\tau_{f} \sim 1000$ y) 
through gravitational instability \citep{B05}, 
or slowly through core accretion \citep{L01} where agglomeration of dust grains creates a rocky core that accretes surrounding gas.  
Core accretion has been estimated to 
take  $\sim 10$ My, although recent 
arguments \citep{r04,mp05} lowering this 
to $\sim  $Myr may be required by 
SST observations.

SST has provided a rapidly growing database of high resolution IR spectral observations of young 
disk-star systems.   From the deficit of emission at wavelengths characteristic of the disk
inner regions, the observations imply that some systems 
have mass-depleted inner disks:
Surrounding the star CoKuTau/4 is a disk with a $9$ AU spectral hole bounded by a very sharp outer 
edge, strongly indicative of the influence of a planet. 
Within 9 AU, the hole seems to require a density to be $\le 10^{-4}$ times that which would otherwise be present if the disk inferred at 9AU
were extended via standard models all the way to the star (D'Alessio et al. 2005).
Such holes extend 
are beyond the scale for which a magnetic field would be important \citep{k91}
and the sharpness is unlikely to be explained by a  radiation pressure or a wind \citep{2005ApJ...621..461D}.
While the presence of a planet is currently the 
most plausible explanation for the sharp edge and hole, the young age \citep{2005ApJ...621..461D}  
of the system, $\leq 2 $My
presents challenges for planet formation models \citep{Q04}.  SST observations of other systems reveal 
direct evidence for accretion in systems with and without 
spectral holes (Muzerolle et al., 2005, N. Calvet \& D. Watson, personal communication)

Here we show, via direct numerical simulations, that
Jupiter mass planets orbiting a solar mass star 
clear out inner holes all the way to the star 
on a time scale faster than the viscous or planet migration
time scales. {\refe We identify this fast accretion with
enhanced angular momentum transport by spiral waves induced by the planet
and we find that a specific limit of the 
general theory of Takeuchi et al. (1996)
of spiral wave angular momentum transport in turbulent disks 
offers a consistent explanation of our results.

We emphasize that in our calculations, the planet is embeded in an already 
turbulent disk with an $\alpha$-viscosity \cite{SS73}.
Our approach therefore differs from e.g \cite{S87},
for which the disk has no means of angular momentum transport other than
a  dissipation of spiral waves themselves. For the latter, 
\cite{L90} showed that, in the absence of 
turbulent viscous damping, waves will become weak shocks, 
which  can provide 
a  modest effective viscosity \cite{SPL94}.
This mechanism was shown by \cite{S87} to be effective for hot disks
and has been called "wave accretion"  \citep{L90}.
It was originally presented, in the case of white dwarf binaries, as a possible explanation for the 
observed optical periodicities in \cite{MAFW99}. 
While our work here does appeal to 
the impact of spiral waves in the transport of angular momentum, 
it is important that for us these waves  act 
in addition to the baseline disk turbulence.  
In the case considered herein, the pre-existing turbulent
viscosity keeps the waves linear, preventing steepening into shocks 
and thereby allowing transport over farther distances from the wave launch.
point. This increases the effective angular momentum transport from 
the case where weak shocks form as long as the initial viscosity is not too
large.

In addition to showing that our results are consistent with the
theory of Takeuchi et al. 1996, we use the results  to account for the recent 
spectral holes observed in young YSO system  such 
CoKuTau 4 \citep{2005ApJ...621..461D}.}

In section 2 we discuss our numerical calculations of the surface density
and accretion rate evolution. We discuss the results, 
interpretation, and implications in section 3 and  conclude in section 4.

\section{Surface Density and Accretion Rate  Evolution}

To determine whether planets can produce sharp extended spectral holes 
and to calculate the associated mass accretion rates, 
we have carried out a series of 
2-D simulations of planets opening up disk gaps which then evolve 
into holes as material drains onto the star.  While there have been previous numerical simulations
of gap formation \citep{bryden,kley,nelson,V04}, 
our work differs in that we have followed the 
disk accretion significantly longer, capturing 
the inner disk clearing all the way to the star.

\subsection{Numerical Simulation Set-up}

We use a  code 
which eliminates the azimuthal velocity from the computation of the 
Courant-Friedrich-L\'evy condition. \citep{m00,m02,mp03}.  This speeds up the computation and
facilitates studying the long term disk evolution.
Our simulations begin with a viscosity parameter $\alpha = \nu/c h=0.00625$,
where $\nu$ is the viscosity and $c(r)$ is the local sound speed.
The disk height-to-radius ratio $h/r = 0.04$, independent of radius. 
This corresponds to a Reynolds 
number ${\cal R}  \equiv \Omega r^{2}/\nu  = \Omega r^{2}/(\alpha c h) = (1/\alpha) (r/h)^{2} =10^{5}$, where $\Omega(r)$ is the Keplerian speed.
We performed simulations with $100 \times 300$ grid cells 
and $150 \times 450$ grid cells to test for convergence.

{\refe In the absence of a planet, accretion at any radius occurs
at the viscous time $t_\nu={\cal R}/\Omega(r)$  ($\sim 1.6\times 10^{4}$ yr at $1$AU). Here were compare this to the case
when a planet is present.}
We ran simulations for a variety of planet masses $M_p= 0.001, 0.002$, and $0.005M_\odot$, 
orbital radii $r_p= 1, 5$ and $7$ AU, and initial surface density profiles $\Sigma \propto r^{-q}$ with $q=0,1$. The two values for $q$
 exhibit similar results. We focus on the $q=1$ case herein. The planet was always initially set onto a circular orbit and was free to migrate via 
angular momentum exchange with the disk.  The planet was 
allowed to accrete gas within its Roche Lobe.
Self-gravity is unimportant for our disks.
 
\subsection{Quantitative Results}

Fig. 1 shows $\Sigma(r)$ at five different times for a case in which 
$r_p = 1$ AU and $M_p=0.001\ M_\odot = 1\ M_{J}$.  
As seen at $t=1000$,  the surface density in the inner regions first increases as matter is pushed 
toward the star and the gap opens, but  an inner hole forms subsequently 
as material is lost to the star.  
The gap near 1AU  forms in less than $1000$ orbits and a hole cleared
all the way to the star  
(i.e. $\Sigma(r_*\le r\le r_p)$ drops by factor $\ge 10$ for all
$r_*\le r\le r_p$, where $r_*$ is the
radius of the star) clears by $\sim 2000$ orbits 
which is $\sim 1/10$ the viscous time at $r=r_p$, 
By 6000 orbits, the surface density has droped by a nearly a factor of 
$10^{-5}$.at 1AU and by $10^{-3}$ at 0.4AU.

Fig. 2  shows  the time evolution of disk  mass $M_r(t)$ 
within $r_{\star} \le r \le r_p$.  The cases shown have $M_p=0.001\ M_{\odot} = 1\ M_{J}$  
and $M_p=0.002\  M_{\odot}= 2\ M_{J}$  and $r_p = 1$ AU.  
From $ 900 < t < 2000$ orbits, $M_r(t)$ rapidly 
depletes by a factor of $25$. This is $\sim 0.1$ viscous accretion
times at $r=r_p$ and consistent with the rapid gap formation,
followed by a slower, but still faster-than-viscous 
depletion at smaller $r$.
{\refe The upper, almost straight, line in Fig. 2 represents  the mass evolution in the
case with  NO planet in a disk limited to a $1$ AU size,
for comparison with the much faster time evolution of the surface
density when the planet is present.  The distiction shows that the enhanced
surface density evolution is not a boundary condition effect.}

Fig. 3 shows the accretion rate  onto to the star, ${\dot M}(t)$, 
for the two mass cases.  
Measurements of SEDs for disks are sensitive to dust only, and the gas density is inferred from 
assuming a gas-to-dust density ratio \citep{2005ApJ...621..461D}.
Direct measure of $\rho_g$ requires detection of 
molecular lines such as those from CO, which is difficult.  For a given $\alpha$, measuring $\dot M$ provides an alternative 
measure of $\rho_{g}$.  Observations of proto-planetary disks with SEDs showing 
holes often indicate accretion rates 
${\dot M}\sim 10^{-9}{M}_\odot$/yr.  \citep{muzerolle05}.  Despite the spectral holes, 
this rate is still large, of order $10\%$ of the rates typical for similar disks without an inner hole  
(called Class II objects). 
The accretion rate at any given time depends on 
the initial disk mass, which can vary between systems, so an important quantity to extract from simulations 
is the ratio of the initial $\dot M$ to that after the spectral hole is present.  
In our simulations, 
the initial parameters give $\dot{M}(t=0) \simeq 4\times 10^{-8} \ M_{\odot}/y$. Focusing on the $M_p=1 M_{J}$  case of Fig. 3, $\dot M$ rises 
to nearly ${\dot M}=100\dot{M}_a(t=0)$ at $t=1900$y, 
and subsequently drops 
to ${\dot M}\sim 0.05\dot{M}_{a}(t=0)$ at $t = 6000$ y 
after which it continues to decay. 
Figs. 1-3 thus demonstrate both (1) rapid  hole 
formation and  (2) substantial accretion rates even after
a hole of with a $10^{-3}$ deficit in surface density has formed.
Note also that the absence of significant planet migration on the hole clearing time
suggests that spectral holes likely  accompany  massive planets.
 
{\refe  In fig \ref{fig:sed} we compare the SED at two different times during the hole formation
with the planet-free case. The figure shows the planet's effect on the spectra as early as $3000$ orbits 
(on the left). The effect  grows stronger with time until a spectral hole forms.}

\section{Interpretation and Implications}

\subsection{The role of spiral waves in clearing the gap and hole}

The faster-than-viscous rapid hole formation exhibited in Figs.1-3
is due to outward angular 
momentum transport at $r<r_p$ from spiral waves driven off of resonances 
between the orbital motion of the planet and  disk material \citep{gt80,takeuchi}. 
{\refe As mentioned in the introduction,  
our approach is different from that of \cite{S87,L90} and \cite{MAFW99},
as we take an already viscous disk. 
\cite{SPL94} 
showed that, in the absence of 
viscous damping,  weak shocks produced by the spiral waves, 
provide an effective viscosity $\alpha \sim (c/r\Omega)^{3})$.  
We therefore proceed to analyze our results in the context of Takeuchi et al. (1996) who considered spiral wave induced angular momentum
transport in which turbulent viscosity damps the waves sufficiently to sustain
keep them linear, before
they steepen weak shocks. Since angular momentum is transported
only over the damping length, a finite viscosity allows
waves to transport angular momentum over a large distance.
If the initial viscosity
were too high, then the waves would have no effect. 
Our viscosity and planet-to-star mass ratio satisfy
the condition in Takeuchi et al. 1996 for enahnced transport
to be expected.}
 
The time evolution of the surface density $\Sigma(r)$ satisfies
\begin{equation}
\frac{\partial \Sigma}{\partial t} = \frac{3}{r}\frac{\partial}{\partial r}\left[
r^{1/2} \frac{\partial}{\partial r} \left(\nu \Sigma r^{1/2}\right) -
\frac{r^{1/2}}{2\pi (GM_{\star})^{1/2}}T\right],
\end{equation}
where the first term on the right is due to viscous accretion and the second is due to the action 
of spiral waves.  The total torque per unit length produced by the planet on the disk is given by    
$T =\sum_{m} T_{m} \propto \pm M_{p}^{2} \Sigma r^{4} /(M_{\star} (r-r_{p})^{4}) $, 
where the subscript m refers to the m-th orbital Lindblad resonance
 \citep{gt80,takeuchi}.   
Interior 
to the planet, spiral waves enhance the 
outward angular momentum transport from a purely viscous disk.  
They transport angular momentum out to a distance to which the
waves damp. A formula for the damping length $l$ is given by
\cite{takeuchi} as
\begin{equation}
\left[\left(\zeta +\left({4\over 3}+{\kappa^2\over m^2 (\Omega-\Omega_p)^2}\right)\right)\nu{m(\Omega_p-\Omega)\over c^2}kl\right]_{r_L-l}\simeq 1,
\label{td0}
\end{equation}
where $\kappa\simeq \Omega$ is the epicyclic frequency, $\nu$ is the shear viscosity,
$\zeta$ is the bulk viscosity, $k$ is radial wavenumber of the mode,
and $m$ is the azimuthal wave number, 
and $r=r_L-l$ is the radius at which quantities are to be evaluated,
where $r_L$ represents the location of the Lindblad resonance
at which the wave of mode $m$ is launched.

Near $r=r_p$, 
the angular momentum transport is 
initially dominated by the second term on the right of Eq. (1)  from waves launched at the $m \sim r/h  \simeq 25$ resonance \citep{takeuchi}. Assuming that
the initial gap corresponds to the damping length of these large $m$ 
waves, Takeuchi et al. (1996) obtain for the gap width
(\ref{td0})
\begin{equation}
\Delta r\sim l\sim r_p(c/r\Omega)_p \alpha^{-2/5}\sim (r_p/3) (h/r/0.04)(\alpha/0.006)^{-2/5},
\end{equation}
where $\alpha$ is specifically used to replace the combination of
$\zeta + 4\nu/3$.
The $m\sim r/h$ waves produce the gap seen near $r=r_p$ at $1000$ orbits in Fig 1.

Although the 
large $m$ linear waves that form the initial gap 
transport angular momentum faster than small $m$ waves, the latter 
damp over a shorter distance. Once large $m$ waves 
clear out a gap, the decrease in surface density at their 
launch location reduces their subsequent contribution to the global angular
momentum transport, and the influence of  
the higher density inner regions takes over.
We now argue that the faster-than-viscous hole clearing by
$t\sim 2000$  is consistent with low $m=2$ waves launched from their
Lindblad resonance at (e.g. \cite{Shu92}) $r_L=(1-1/m)^{2/3}\sim 0.63r_p$ 
dominating  $T$. (We note that $m=2$ are the lowest modes
relevant for a Kelperian disk but $m=1$ could be relevant for sub-Keplerian
disks for which $\Omega$ differs significantly from   $\kappa$.)

The damping length $l$ in the limit of  low $m$ waves is not
explicitly estimated in \cite{takeuchi}, but 
we can estimate it from the low $m$ limit of (\ref{td0}).
To do so, we assume (to be justified later)  that 
the damping length $l$ is large enough that $r_L -l << r_p$, so 
$\Omega(r_L-l)-\Omega_p \sim \Omega (r_L-l)$. 
Then (\ref{td0}) becomes
\begin{equation}
(\zeta + (4/3 + 1/m^2) \nu) (m k l\Omega/c^2) \simeq 1.
\label{td1}
\end{equation}
We use the approximation 
$\zeta + (4/3 + 1/m^2)\nu \sim \zeta + 4/3\nu \sim {\alpha} c h.$
so that (\ref{td1}) becomes 
\begin{equation}
l = 1/(\alpha k m), 
\label{td3}
\end{equation}
where (\cite{takeuchi})
\begin{equation}
k=\left[{m^2(\Omega-\Omega_p)^2-\kappa^2\over c^2}\right]^{1/2}
\sim {(m^2-1)^{1/2}\over h(r_L-l)}\sim \left({3^{1/2}\over 0.63r_p-l}\right){r\over h}  
\label{td4},
\end{equation}
and the latter similarity follows for 
$r_L=0.63r_p$  and  $m=2$. 
Using (\ref{td4}) in  (\ref{td3}) then gives
\begin{equation}
l =  {0.36r_p}{h\over r}{1\over \alpha }\left(1+
{0.5\over 3^{1/2}}{h\over r}{1\over \alpha }\right)^{-1}.
\label{td6}
\end{equation}
If we take $\alpha = 0.006$ and $h/r= 0.04$ as per our simulations, then
for $m=2$, $l \sim 0.42 r_p$, so that $r_L-l=0.63r_p-l \sim 0.21r_p$.
The viscous time at $r=0.21r_p$ is ${{\cal R}\over 2\pi}(0.21)^{3/2}{\rm yr}
\sim 1530$yr.  Fig. 1 shows that by $2000$yr, $\Sigma(r)$ has decreased
by at least a factor of 10 from its initial value for all radii.
Were this purely the result of viscous evolution, we would
see such clearing only for $r\le 0.2r_p$.  In short, the clearing at all radii
out to $r_p$ illustrates the initial 
influence of $m=r/h\sim 25$ spiral waves followed
by  the subsequent influence of $m=2$ spiral waves.

{\refe 

We can define the hole formation time 
as the time scale at which the surface
density at $r\le 2r_p/3$ drops by factor of 10.
We use this radius since, as discussed above, 
 the gap opens from $r_p\ge r \ge r\sim 2r_p/3$.
Thus the hole formation represents the remaining clearing
inside $r\le 2r_p/3$. 
The mass scaling for the hole clearing time scale 
can be estimated  using $\tau_h \sim {\Sigma \over \partial_t \Sigma}$
and the third term of Eq. (1).
Measured in units of the planet's orbit period,
the result is 
\begin{equation}
\tau_{h}(2r_p/3)\Omega(r_p) \sim {0.1}{{\cal R}\over 2\pi}
\left({M_*\over 1M_\odot}\right)^{3/2}\left({M_p\over 1M_J}\right)^{-2},
\label{hole}
\end{equation}
where the numerical coefficient $0.1$ comes from the simulations (e.g. Fig. 1) while the parameter scalings come from the theory.
The result (\ref{hole}) is $\sim 0.16$  the viscous time at $r=2r_p/3$
and $0.1$ times the viscous time at $r=r_p$.}

\subsection{Implications for observing holes and gaps}

Finally, we point out  that for a planet to have any observable effect,
it must form faster than the migration time scale 
at its formation location. 
In our planet-disk simulations, we imposed 
a planet at $t=0$, so the planet formation time scale 
is unspecified. However, a constraint emerges.
If the planet growth time is long compared to the hole formation
time scale, $\tau_h$, then a spectral hole should always
accompany the planet.  If instead the planet formation time, $\tau_f<\tau_h$, 
then the fraction of systems with a planet but without a hole,
would be  
$f\sim {\tau_{h}(2r_p/3)\over \tau_{mig}}$, 
and $\tau_{mig}$ is the migration time at the planet's inferred location.
From the simulations, we find that  
 $\tau_{mig}\Omega(r_p)\ge {\cal R}/2\pi$ 
(where the equality corresponds to $\Sigma(r,t=0)\propto r^{-q}$ with $q\ge 0$)
. 
{\refe 
Then, using (\ref{hole})
$f \sim {0.1}\left({M_*\over 1M_\odot}\right)^{3/2}\left({M_p\over 1M_J}\right)^{-2}
$. }

Our key result, that massive planets rapidly form spectral holes in disks, 
thus implies: 
(1) For planet formation on a time scale $\tau_{f}>\tau_h$, 
all disks with a massive planet will have a sharply bounded 
hole  from outer radius $r_{out}=r_p$ to inner radius $r_{in}=r_*$.
(2) For $\tau_{f}<\tau_h$, a  fraction $f$ of disks with a massive 
planet would show a sharply bounded gap from $r_{out}> r_p$ to $r_{in}>r_*$.
(3) However, since $f$ is small for massive planets. Therefore, 
disks which appear to show gaps but not holes \citep{mm92,mm93} 
are best explained either by planets with low enough mass to make 
$\tau_h$ as long as the viscous time (in which case only the early 
gap formation is observed), or by spectral features from dust and inclination
effects without a planet.

\section{Conlcusion}

We find that a massive planet
embedded in a  protoplanetary accretion disk 
can produce a  low surface density hole faster than
can be accounted for by purely viscous evolution.
In particular,
(1)  after less than $1000$ orbits 
 a very low 
surface density gap forms that extends  approximately 1/3 the distance
to the star.  (2) By about  $2000$
orbits,
a low surface density hole, with a factor of $\ge 10$ density depletion
extending all  the way from the planet orbital radius
to the stellar surface appears on a time scale $\sim 0.16$ of the 
 viscous evolution time calculated at $r=2r_p/3$.

The  gap and hole are consistently interpreted to be 
 the result of the enhanced 
angular momentum transport and accretion induced by  spiral waves
\citep{gt80,artymowicz94,takeuchi}. 
We find that the gap can be explained by the launching and damping
of of $m=r/h\sim 25$ waves and the faster-than-viscous hole formation
is consistent with launching and damping of $m=2$ waves.
Our results are consistent with previous studies of gap formation and
the theory of Takeuchi et al. (1996) but 
we have followed the evolution of the disk long much longer than in
previous simulations \citep{bryden,kley,nelson,V04}
and we are able to see the hole
clear all the way to the star. 

Our results imply that spectral holes could be more 
common than spectral gaps  when a sufficiently 
 massive planet is present inside
a disk. We also emphasize that the use of the 
term ``spectral hole'' is important because
even though the holes have a reduced density for all $r_*< r< r_p$, 
they need  not be fully evacuated and 
substantial accretion rates can be maintained even after the hole forms.

{\bf Acknowledgments}:

We thank N. Calvet, W. Forrest, and D. Watson for helpful discussions. 
PV Thanks F.Masset for the code used here.
We acknowledge support from NSF grants AST-0406799, AST
00-98442, AST-0406823, NASA grant ATP04-0000-0016, and the KITP of UCSB, where this research was supported in part by NSF Grant PHY-9907949.  We also acknowledge support from the Laboratory for Laser Energetics.

\clearpage

\begin{figure}
\centerline{\epsfig{file=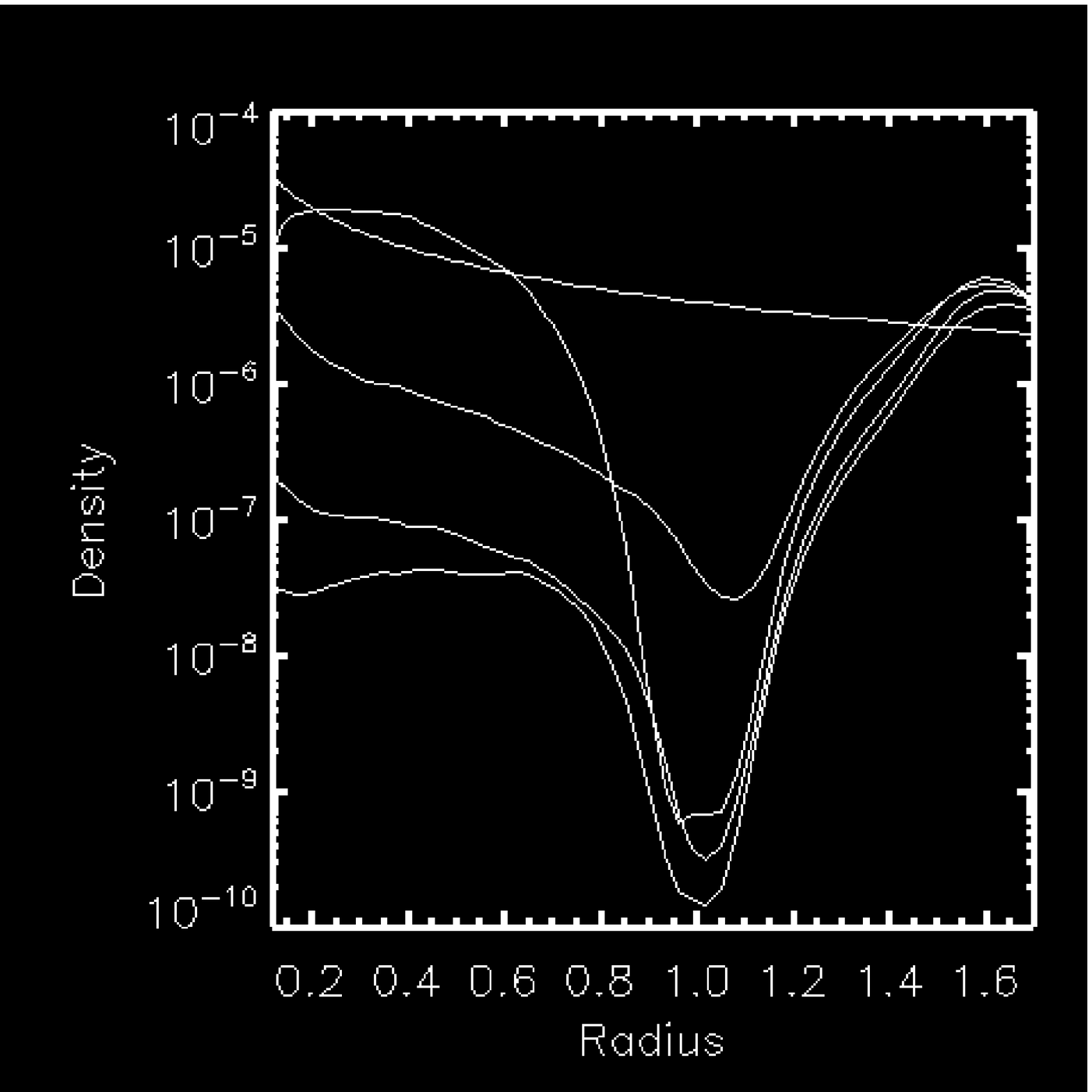,width=\linewidth}}
\caption{Evolution of the density in the inner region of the disk. 
There is a $1\ M_{J}$ planet at $1$AU. The different curves
(identified from the top down at the inner edge)   
are $0$, $1000$, $2000$, $4000$, and $6000$ orbital times at 1AU.}
\label{fig:density_evolution}
\end{figure} 

\clearpage

\begin{figure}
\centerline{\epsfig{file=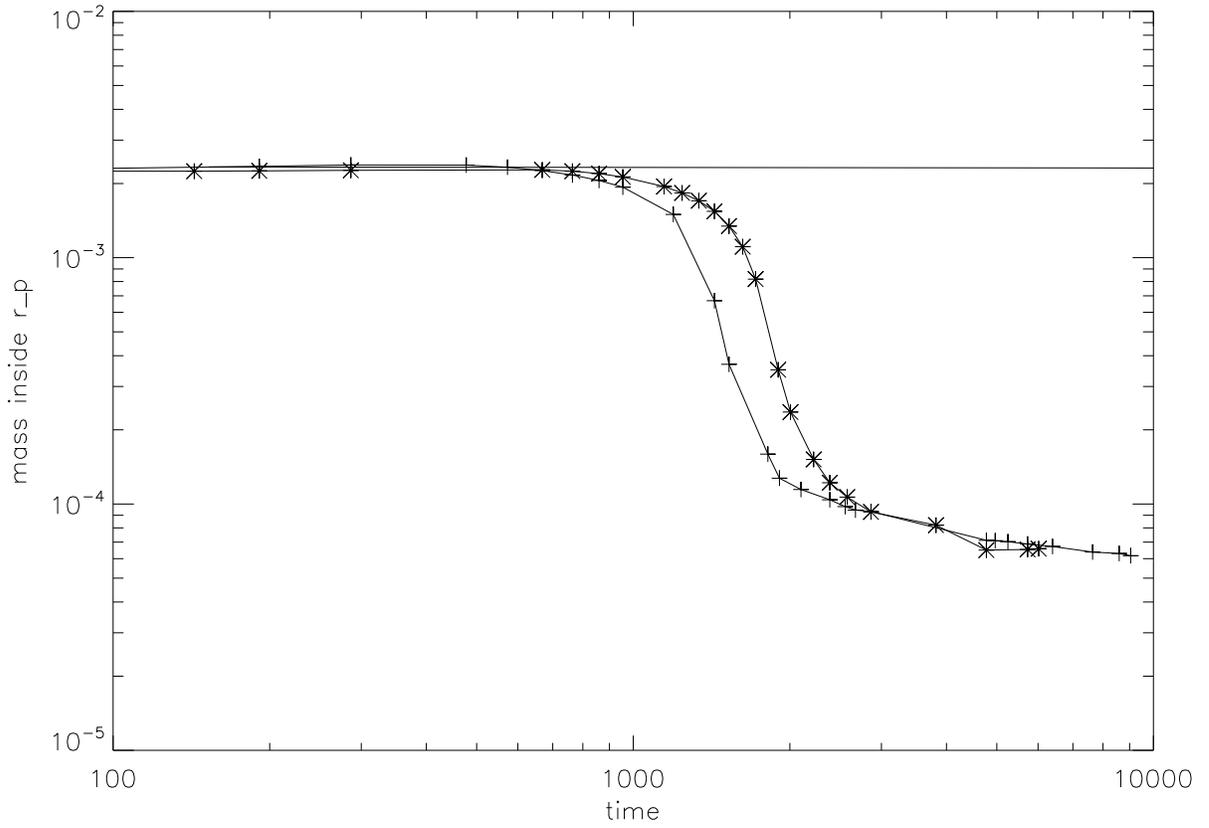,width=\linewidth}}
\caption{Evolution of the mass interior to the planet as a 
function of time for a planet at $1$AU.
the ''*'' represents a $1\ M_{J}$ planet and the ''+'' a $2\ M_{J}$ planet.
The almost straight line is the same plot in the case of no planet in the system but
a disk size limited to $1$AU.}
\label{fig:mass_evolution}
\end{figure}

\clearpage

\begin{figure}
\centerline{\epsfig{file=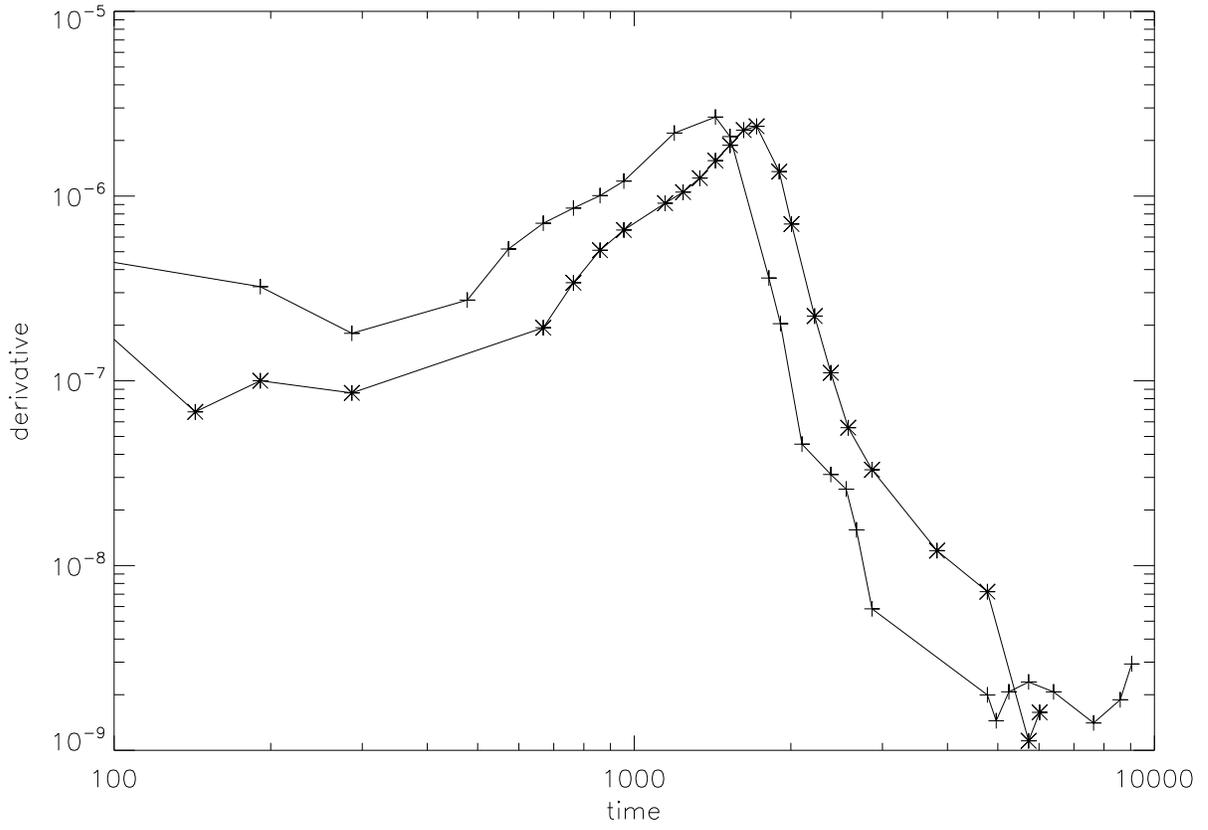,width=\linewidth}}
\caption{Evolution of the mass accretion rate as function of time for a planet at $1$AU.
the ''*'' represents a $1\ M_{J}$ planet and the ''+'' a $2\ M_{J}$ planet.}
\label{fig:mass_accretion_rate_evolution}
\end{figure}

\clearpage

\begin{figure}
\centerline{\epsfig{file=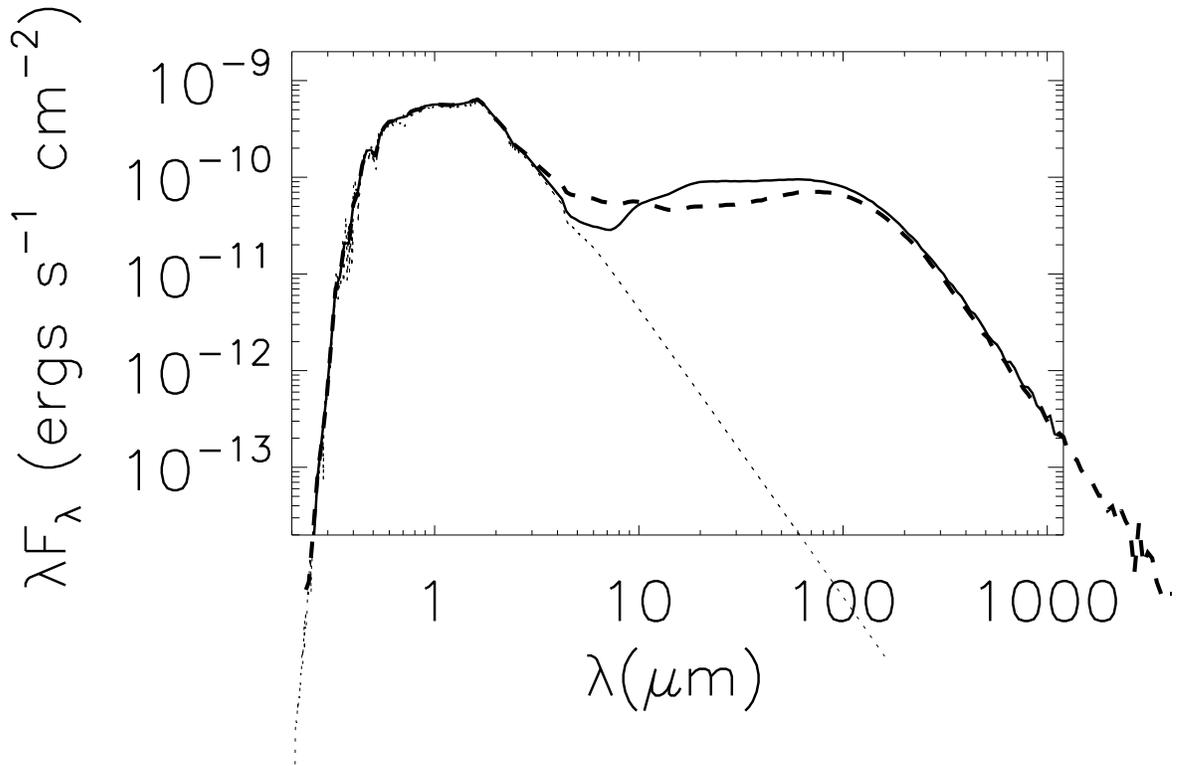,width=\linewidth}}
\caption{Spectral Energy Distribution (SED) after $10000$ orbits,
compared with the no-planet case. We see the creation of a spectral hole.}
\label{fig:sed}
\end{figure}

\end{document}